%% file: main.tex
\documentclass[10pt,conference]{IEEEtran}
\usepackage{iftex}
\ifXeTeX 
  \usepackage{fontspec}
\else
  \usepackage[T1]{fontenc} 
\fi
\usepackage{booktabs}
\usepackage{array}
\newcolumntype{L}[1]{>{\raggedright\arraybackslash}p{#1}}
\usepackage{tabularx}
\newcolumntype{Y}{>{\raggedright\arraybackslash}X}
\usepackage{xcolor}
\usepackage{tikz}
\usetikzlibrary{arrows.meta,positioning,shapes.multipart,calc}
\usepackage{url}
\usepackage{cite}
\usepackage{xspace}
\usepackage[hidelinks]{hyperref}
\hypersetup{pdftitle={RootSink: Reproducible C/C++ Crashes with Tested Repair Sites},
  pdfauthor={Suryansh Singh Sijwali, Achintya Lakshmanan, Abhiram Dodda, Shaiful Chowdhury, Suman Saha}}
\input{generated/macros.tex}
\newcommand{\CF}{RootSink\xspace}
\newcommand{\cwe}[1]{\mbox{CWE-#1}}

\newcommand{\para}[1]{\textbf{#1:}\ }
\newcommand{\paraq}[1]{\textbf{#1}\ } 

\begin{document}

\title{\CF: Reproducible C/C++ Crashes with Tested Repair Sites}

\author{
  \IEEEauthorblockN{
    Suryansh Singh Sijwali\IEEEauthorrefmark{1},
    Achintya Lakshmanan\IEEEauthorrefmark{1},
    Abhiram Dodda\IEEEauthorrefmark{1},\\
    Shaiful Chowdhury\IEEEauthorrefmark{2},
    and Suman Saha\IEEEauthorrefmark{1}
  }
  \IEEEauthorblockA{
    \IEEEauthorrefmark{1}The Pennsylvania State University\\
    University Park, PA, USA\\
    \{sss6371, aql6062, amd8734, szs339\}@psu.edu
  }
  \IEEEauthorblockA{
    \IEEEauthorrefmark{2}University of Manitoba\\
    Winnipeg, MB, Canada\\
    Shaiful.Chowdhury@umanitoba.ca
  }
}

\maketitle

\input{sections/abstract}
\input{sections/introduction}

\input{sections/background}
\input{figures/pipeline}
\input{sections/building}
\input{sections/dataset}

\input{tables/related}
\input{sections/validation}
\input{sections/related}
\input{sections/limitations}
\input{sections/availability}

\bibliographystyle{IEEEtran}
\bibliography{refs}

\end{document}

%% file: generated/macros.tex
\newcommand{\NRecords}{3341\xspace}
\newcommand{\NGold}{1954\xspace}
\newcommand{\NGoldUpstream}{410\xspace}
\newcommand{\NSilver}{751\xspace}
\newcommand{\NBronze}{226\xspace}
\newcommand{\NUpstreamRan}{195\xspace}
\newcommand{\NUpstreamNoRepro}{142\xspace}

\newcommand{\NFixStillCrashes}{15\xspace}
\newcommand{\NProjects}{354\xspace}
\newcommand{\NAdvisoryOnly}{428\xspace}
\newcommand{\NRecArvo}{2241\xspace}
\newcommand{\NRecSyzbot}{672\xspace}

\newcommand{\DatasetMB}{37\,MB\xspace}
\newcommand{\LogsGB}{1.5\,GB\xspace}
\newcommand{\NProjectsGold}{226\xspace}

\newcommand{\NGoldArvo}{1763\xspace}

\newcommand{\PctGoldArvo}{90\%\xspace}

\newcommand{\NGoldBugs}{1746\xspace}
\newcommand{\NGoldLenThree}{67\xspace}

\newcommand{\NGoldProven}{754\xspace}

\newcommand{\PctFixAwayFromCrash}{50\%\xspace}
\newcommand{\NFixAwayBase}{1954\xspace}
\newcommand{\PctFixOffStackArvo}{14\%\xspace}
\newcommand{\PctFixOffCrashStackArvo}{21\%\xspace}
\newcommand{\NFixStackBaseArvo}{1763\xspace}

\newcommand{\NGoldDated}{1898\xspace}
\newcommand{\NGoldAfterTwentyFive}{118\xspace}
\newcommand{\NGoldUntestedRoot}{679\xspace}

\newcommand{\PctGoldUntestedRoot}{35\%\xspace}

\newcommand{\NGoldRootExec}{748\xspace}
\newcommand{\NGoldRootAttested}{69\xspace}
\newcommand{\NGoldRootInferred}{362\xspace}
\newcommand{\NGoldRootDataflow}{19\xspace}
\newcommand{\NGoldRootNone}{2\xspace}

\newcommand{\NGoldTestedRoot}{1275\xspace}
\newcommand{\NRootSiteTested}{1273\xspace}
\newcommand{\NRootSiteOrigin}{434\xspace}
\newcommand{\NRootSiteAnnotated}{70\xspace}
\newcommand{\NRootSiteNone}{177\xspace}

\newcommand{\NGoldTrain}{1333\xspace}
\newcommand{\NGoldValidation}{306\xspace}
\newcommand{\NGoldTest}{315\xspace}

\newcommand{\NShownTested}{81\xspace}
\newcommand{\NShownOrigin}{411\xspace}

\newcommand{\NShownTestedFn}{843\xspace}
\newcommand{\NGoldRootFormerTwenty}{995\xspace}

\newcommand{\NGoldRootClassLevel}{425\xspace}

\newcommand{\NGoldFixAttr}{1429\xspace}
\newcommand{\NGoldNotFixAttr}{123\xspace}
\newcommand{\NGoldFixAttrUnknown}{402\xspace}

\newcommand{\NCoreFixExec}{1128\xspace}
\newcommand{\NCoreProjects}{183\xspace}
\newcommand{\NTestedCrashLine}{234\xspace}
\newcommand{\NTestedCrashFn}{209\xspace}
\newcommand{\NTestedOnStack}{306\xspace}
\newcommand{\NTestedOffStack}{413\xspace}
\newcommand{\NTestedRelKnown}{1162\xspace}

\newcommand{\BaseRootOld}{98\%\xspace}
\newcommand{\BaseRoot}{58\%\xspace}

\newcommand{\BaseLocN}{208\xspace}
\newcommand{\BaseLocFn}{37\%\xspace}

\newcommand{\BaseLocStack}{67\%\xspace}

\newcommand{\NPbGold}{30\xspace}
\newcommand{\NPbTested}{22\xspace}
\newcommand{\NPbTestedFn}{8\xspace}
\newcommand{\NPbTestedFile}{6\xspace}
\newcommand{\NPbOrigin}{6\xspace}

\newcommand{\NPbTestedOther}{8\xspace}
\newcommand{\NPbAnnotated}{2\xspace}

\newcommand{\ArvoInputPhrase}{every ARVO record\xspace}
\newcommand{\NChainUninit}{441\xspace}
\newcommand{\NChainUafRead}{344\xspace}
\newcommand{\NChainCheckRead}{750\xspace}
\newcommand{\NChainCheckWrite}{310\xspace}
\newcommand{\ExHunks}{eight\xspace}
\newcommand{\ExRootCwe}{\cwe{1284}\xspace}
\newcommand{\ExSinkCwe}{\cwe{125}\xspace}
\newcommand{\NGoldTestedOneHunk}{749\xspace}
\newcommand{\PctGoldTestedOneHunk}{59\%\xspace}
\newcommand{\DataDOI}{\url{https://doi.org/10.5281/zenodo.24068722}\xspace}

\newcommand{\ReleaseVersion}{1.0.10\xspace}

\newcommand{\IrrSinkCwe}{97\% (95\% CI 92--99\%)\xspace}

\newcommand{\IrrSinkSite}{96\%\xspace}
\newcommand{\IrrSrcCwe}{64\% (95\% CI 55--72\%)\xspace}
\newcommand{\IrrSrcCwePct}{64\%\xspace}
\newcommand{\IrrSrcHier}{79\%\xspace}
\newcommand{\IrrSrcHierCur}{72\%\xspace}

\newcommand{\NIrrTested}{53\xspace}
\newcommand{\IrrStrictTested}{68\% (95\% CI 55--79\%)\xspace}
\newcommand{\IrrPctStrictTested}{68\%\xspace}
\newcommand{\IrrStrictTestedCI}{55--79\%\xspace}

\newcommand{\NIrrOrigin}{51\xspace}
\newcommand{\IrrStrictOrigin}{8\% (95\% CI 3--18\%)\xspace}
\newcommand{\IrrPctStrictOrigin}{8\%\xspace}
\newcommand{\IrrStrictOriginCI}{3--18\%\xspace}

\newcommand{\NIrrAnnot}{13\xspace}

\newcommand{\NIrrStrictAnnot}{0\xspace}

\newcommand{\NIrrOriginCwe}{50\xspace}
\newcommand{\NIrrOriginInFixFn}{24\xspace}

\newcommand{\NIrrTestedCwe}{26\xspace}

\newcommand{\NIrrMoved}{30\xspace}
\newcommand{\NIrrMovedDiffer}{27\xspace}
\newcommand{\NIrrMovedTestedFn}{25\xspace}

\newcommand{\NIrrMovedOriginOnly}{0\xspace}
\newcommand{\NIrrMovedViaCrash}{6\xspace}

\newcommand{\IrrPooledStrict}{37\%\xspace}

\newcommand{\IrrBSrcCwePct}{70\%\xspace}
\newcommand{\IrrBSrcHier}{84\%\xspace}
\newcommand{\IrrBSrcHierCur}{78\%\xspace}

\newcommand{\IrrBPctStrictTested}{92\%\xspace}
\newcommand{\IrrBStrictTestedCI}{82--97\%\xspace}

\newcommand{\IrrBPctStrictOrigin}{14\%\xspace}
\newcommand{\IrrBStrictOriginCI}{7--26\%\xspace}

\newcommand{\NIrrBOriginInFixFn}{21\xspace}

\newcommand{\NIrrBTestedCwe}{32\xspace}

\newcommand{\NIrrBMovedTestedFn}{21\xspace}

\newcommand{\NIrrBMovedOriginOnly}{1\xspace}
\newcommand{\NIrrBMovedViaCrash}{6\xspace}

\newcommand{\IrrBPooledStrict}{51\%\xspace}

\newcommand{\NIrrPopTested}{708\xspace}
\newcommand{\NIrrPopOrigin}{691\xspace}
\newcommand{\NIrrPopAnnot}{59\xspace}
\newcommand{\NIrrPop}{1458\xspace}
\newcommand{\NIrrLocated}{117\xspace}
\newcommand{\IrrOriginCwePair}{50 of 51 cases each\xspace}
\newcommand{\IrrSinkCweItems}{97\% of items each\xspace}
\newcommand{\IrrAnnotPair}{none of the 13 suggested sites\xspace}
\newcommand{\IrrN}{150\xspace}
\newcommand{\IrrAgreeSource}{83\%\xspace}
\newcommand{\IrrKappaSource}{0.80\xspace}
\newcommand{\IrrKappaSourceCI}{0.72--0.86\xspace}
\newcommand{\IrrAcOneSource}{0.82\xspace}

\newcommand{\IrrAgreeSite}{75\%\xspace}
\newcommand{\IrrKappaSite}{0.70\xspace}

\newcommand{\IrrAgreeSiteFn}{84\%\xspace}
\newcommand{\NIrrSiteBothUnclear}{17\xspace}

\newcommand{\NAiModels}{3\xspace}
\newcommand{\AiTestedRange}{64--70\%\xspace}
\newcommand{\AiOriginRange}{6--20\%\xspace}

\newcommand{\ResourceFigures}{a scored rebuild took a median of 88\,s and used 0.8\,GB (95th percentiles 20\,min and 11\,GB)\xspace}
\newcommand{\DataURL}{\url{https://github.com/SuryanshSS1011/RootSink}\xspace}

\newcommand{\RerunCleanStillGold}{96\xspace}
\newcommand{\RerunCleanN}{100\xspace}
\newcommand{\RerunCleanPct}{100\%\xspace}
\newcommand{\RerunCleanCI}{96--100\%\xspace}

\newcommand{\DupSpanPhrase}{no duplicate group spans two splits\xspace}

\newcommand{\NRlReadyKernelGold}{173\xspace}
\newcommand{\NKHostedTried}{163\xspace}
\newcommand{\NKHostedRun}{157\xspace}
\newcommand{\NKHostedFired}{147\xspace}

\newcommand{\NRlScoredGold}{1730\xspace}
\newcommand{\NRlReadyGold}{1238\xspace}

\newcommand{\CovTolPct}{90\%\xspace}

\newcommand{\NRlCovGold}{377\xspace}

\newcommand{\NFromSrcN}{60\xspace}
\newcommand{\NFromSrcBuilt}{13\xspace}
\newcommand{\NFromSrcRepro}{11\xspace}

\newcommand{\NFuzzWithdrawn}{46\xspace}
\newcommand{\NFuzzTestsAll}{1145\xspace}

\newcommand{\NFixDistReadyN}{29\xspace}

\newcommand{\NFixDistReadyDiff}{28\xspace}

\newcommand{\NIrrToMapped}{34\xspace}

\newcommand{\NIrrNowTestedOrigin}{33\xspace}
\newcommand{\NIrrNowTestedOther}{9\xspace}
\newcommand{\NIrrDropped}{2\xspace}

\newcommand{\NIrrGold}{130\xspace}
\newcommand{\NIrrOther}{20\xspace}

\newif\ifIrrTwoRaters
\IrrTwoRaterstrue

%% file: sections/abstract.tex
\begin{abstract}
When a C or C++ program crashes on a crafted input, the line that crashes is often not the line
that is wrong. Tools and models that repair or locate vulnerabilities need examples showing both, with evidence for each, which existing datasets rarely provide. \CF provides \NRecords
records from \NProjects open-source projects. Most give the crash site a memory-error detector
reported and a candidate location for its root cause, and label each with a weakness type
and its evidence. For the \NGold records in the Reproduced tier, we re-ran the crash ourselves and confirmed that the fixed version runs cleanly. In \NGoldTestedRoot of these, a single hunk of the developer's
fix stopped the crash on its own. That marks a tested repair site, though not necessarily the defect, and \NRootSiteTested records use it as the root.
\ifIrrTwoRaters In a blind audit of an earlier version, two co-author raters called \IrrPctStrictTested and \IrrBPctStrictTested of the tested repair sites they saw the defect, but only \IrrPooledStrict and \IrrBPooledStrict of that version's located Reproduced-tier roots. \else In a blind audit of an earlier version, still in progress, the first rater judged \IrrPctStrictTested of the tested repair sites seen, but only \IrrPooledStrict of that version's located Reproduced-tier roots, to be the defect. \fi \CF ships with tasks, a patch scorer and run logs.
\end{abstract}

%% file: sections/introduction.tex
\section{Introduction}
Memory-safety bugs in C and C++, such as reading past the end of a buffer, often crash far from
their cause. In Fig.~\ref{fig:example}, from the pjsip library, a STUN network message
declares each attribute's length. The decoder checks that length against the remaining bytes but
forgets the 4-byte header, so a crafted message makes a copy three calls deeper read
past the buffer. The program crashes in the copy, yet the defect is the check.

Repair tools and AI coding agents are usually judged by replaying the crashing input on their patch~\cite{lee2025secbench,mathai2024kgym}. A patch that only guards
the copy passes that test and leaves the defect in place. On bugs fixed away from the crash, this
test inflated the solve rates of 11 agents 1.83-fold on average over stricter checks against input
variants and normal behavior~\cite{shen2026patchbench}.

Training and testing such tools need examples marking both the crash and its cause. Commit-mined datasets label vulnerable functions with advisory weakness types, but they neither locate each weakness nor show that it caused anything~\cite{ni2024megavul,ding2025primevul,croft2023dataquality}. In the widely used BigVul~\cite{fan2020bigvul} and CVEfixes~\cite{bhandari2021cvefixes}, only
25\% and 51.7\% of sampled functions labeled vulnerable were vulnerable~\cite{chen2023diversevul}. Reproducible-crash collections record the crash and its fix~\cite{mei2026arvo}, but few mark the cause, as PatchBench's curated root-cause patches do~\cite{shen2026patchbench}.

\CF narrows this gap with evidence from execution. For each crash it records the crash point, a
candidate location for the cause and the strength of the evidence for each. We re-run each crashing
input we can obtain on both versions and apply each hunk (contiguous block of changed lines) of the developer's fix on its own, since a
hunk that alone stops the crash marks a tested repair site.
\CF contributes (1)~\NRecords records from \NProjects projects, \NGold reproduced by us with every log archived; (2)~candidate root-cause sites with evidence, \NRootSiteTested where one hunk alone stopped the crash, and a blind audit; (3)~three tasks, baselines and a
patch scorer; and (4)~versioned rules with logged corrections.

\input{figures/example}

%% file: figures/example.tex
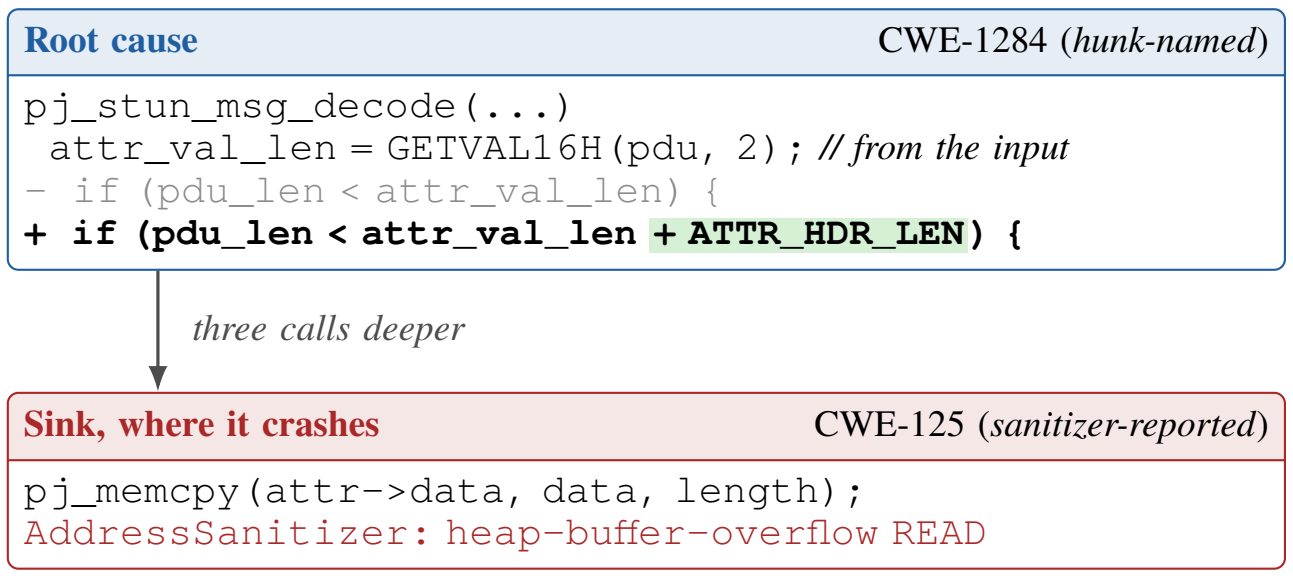
\begin{figure}[t]
\centering
\definecolor{rootc}{RGB}{31,94,160}
\definecolor{sinkc}{RGB}{170,40,40}
\definecolor{addc}{RGB}{214,240,214}
\setlength{\fboxsep}{0.7pt}
\begin{tikzpicture}[font=\scriptsize,
  card/.style={rectangle split, rectangle split parts=2, rounded corners=2pt, draw=#1,
    rectangle split part fill={#1!11, white}, text width=0.93\columnwidth, inner sep=3pt,
    align=left},
  ar/.style={-{Latex[length=1.7mm]}, semithick, black!70}]
\node[card=rootc] (root) {\textcolor{rootc}{\textbf{Root cause}}\hfill
  \ExRootCwe\ (\emph{hunk-named})
  \nodepart{two}\ttfamily pj\_stun\_msg\_decode(...)\\
  \ \ attr\_val\_len = GETVAL16H(pdu, 2); \textrm{\textit{// from the input}}\\
  \textcolor{black!45}{- \ if (pdu\_len < attr\_val\_len) \{}\\
  \textbf{+ \ if (pdu\_len < attr\_val\_len \colorbox{addc}{+ ATTR\_HDR\_LEN}) \{}};
\node[card=sinkc, below=8mm of root] (sink) {\textcolor{sinkc}{\textbf{Sink, where it crashes}}\hfill
  \ExSinkCwe\ (\emph{sanitizer-reported})
  \nodepart{two}\ttfamily pj\_memcpy(attr->data, data, length);\\
  \textcolor{sinkc}{AddressSanitizer: heap-buffer-overflow READ}};
\draw[ar] ([xshift=10mm]root.south west) -- node[right=1mm, font=\scriptsize\itshape] {three calls
  deeper} ([xshift=10mm]sink.north west);
\end{tikzpicture}
\caption{A \CF record from pjsip, simplified. The check omits the \mbox{4-byte} attribute header, which the highlighted term adds, so the copy reads past the buffer. Applied alone, the hunk that adds the bold line, one of the fix's \ExHunks, stops the crash.}
\label{fig:example}
\end{figure}

%% file: sections/background.tex
\section{Background and terms}\label{sec:bg}
\para{Sanitizers} Reading past a buffer or using freed memory (use after free) does not always crash
an ordinary build, so sanitizers catch them with checks the compiler inserts. AddressSanitizer (ASan) catches out-of-bounds and use-after-free
accesses~\cite{serebryany2012asan}, MemorySanitizer (MSan) catches uses of uninitialized
values~\cite{stepanov2015msan}, UndefinedBehaviorSanitizer (UBSan) catches integer
overflows~\cite{ubsan}, and KASAN is the Linux kernel's ASan~\cite{kasan}. A sanitizer reports the error type, call stack and, for ASan and KASAN, whether the access read or wrote.

\para{Sink and root cause} We call the bad access that a sanitizer reports the \emph{sink}, and the
earlier mistake that enables it the \emph{root cause}. In Fig.~\ref{fig:example} the sink is
the read in \texttt{pj\_memcpy} and the root cause is the incomplete check.

\para{CWE} The Common Weakness Enumeration (CWE) is MITRE's public catalog of software weakness
types~\cite{mitre2024cwe}, such as \cwe{125}, an out-of-bounds read. CWE records which weaknesses can
lead to which (\emph{CanPrecede}) and which are special cases of which (\emph{ChildOf}), and it discourages
mapping to broad types such as \cwe{20}, improper input validation. Advisories
in the National Vulnerability Database (NVD)~\cite{nvd} and the Open Source Vulnerabilities database
(OSV)~\cite{osv} often give a bug's CWE and its fixing commit.

\para{Crash sources} A fuzzer feeds generated inputs to a small program, the fuzz target, which
passes each to the code under test, and the fuzzer logs each input it has run. OSS-Fuzz fuzzes many
open-source projects and records each crash~\cite{serebryany2017ossfuzz}. ARVO packages such crashes
as Docker images of the vulnerable and the fixed version, with the crashing input~\cite{mei2026arvo}.
syzbot fuzzes the Linux kernel and often publishes a C reproducer~\cite{syzbot}.

%% file: figures/pipeline.tex
\begin{figure*}[t]
\centering
\definecolor{flowc}{RGB}{31,94,160}
\newcommand{\cnt}[1]{\textcolor{flowc}{\textbf{#1}}}
\begin{tikzpicture}[font=\footnotesize, node distance=4mm,
  box/.style={draw=black!60, rounded corners=2pt, align=center, inner sep=2.5pt},
  source/.style={box, text width=27mm, fill=black!5, minimum height=8.5mm},
  stage/.style={box, minimum height=14mm},
  opt/.style={box, dashed, text width=76mm},
  release/.style={box, text width=23mm, fill=black!5, minimum height=14mm},
  ar/.style={-{Latex[length=1.7mm]}, semithick, black!70}]
\node[source] (syz) {Linux kernel bugs\\ \scriptsize syzbot reports\ \cnt{\NRecSyzbot}};
\node[source, above=1.5mm of syz] (arvo) {OSS-Fuzz crashes\\ \scriptsize ARVO images\ \cnt{\NRecArvo}};
\node[source, below=1.5mm of syz] (osv) {Advisories\\ \scriptsize NVD, OSV\ \cnt{\NAdvisoryOnly}};
\node[stage, text width=25mm, right=8mm of syz] (lab) {\textbf{Label by rule}\\ \scriptsize from the crash report,
  the fix and static paths};
\node[stage, text width=31mm, right=of lab] (rerun) {\textbf{Re-run both versions}\\ \scriptsize vulnerable crashes at
  the sink; fixed runs clean twice};
\node[stage, text width=38mm, right=of rerun] (gate) {\textbf{Assign tier}\\[1pt] \scriptsize
  \begin{tabular}{@{}l@{\ }r@{\quad\ }l@{\ }r@{}}
  Reproduced & \cnt{\NGold} & Upstream & \cnt{\NGoldUpstream}\\
  Partial & \cnt{\NSilver} & Candidate & \cnt{\NBronze}
  \end{tabular}};
\node[release, right=of gate] (rel) {\textbf{Release}\\ \scriptsize \cnt{\NRecords} records, logs,\\ \scriptsize
  task views, scorer};
\node[opt, below=3mm of $(rerun.south)!0.5!(gate.south)$] (abl) {\textbf{Single-hunk test}
  (optional) \scriptsize applies each hunk of the fix alone;\\ \scriptsize the first that stops the crash
  marks a tested repair site};
\coordinate (bus) at ($(syz.east)+(4mm,0)$);
\draw[semithick, black!70] (arvo.east) -| (bus) (osv.east) -| (bus) (syz.east) -- (bus);
\draw[ar] (bus) -- (lab.west);
\draw[ar] (lab) -- (rerun);
\draw[ar] (rerun) -- (gate);
\draw[ar] (gate) -- (rel);
\draw[ar, black!45] (lab.north) -- ++(0,3.5mm) -| (gate.north)
  node[pos=0.25, above, font=\scriptsize\itshape, text=black!60] {advisories seldom have a crash to re-run};
\draw[ar, dashed] (rerun.south) -- (rerun.south |- abl.north);
\draw[ar, dashed] (abl.north -| gate.south) -- (gate.south);
\end{tikzpicture}
\caption{How \CF is built, with records per source and tier. The single-hunk test is optional, and a success adds a tested repair site.}
\label{fig:pipeline}
\end{figure*}
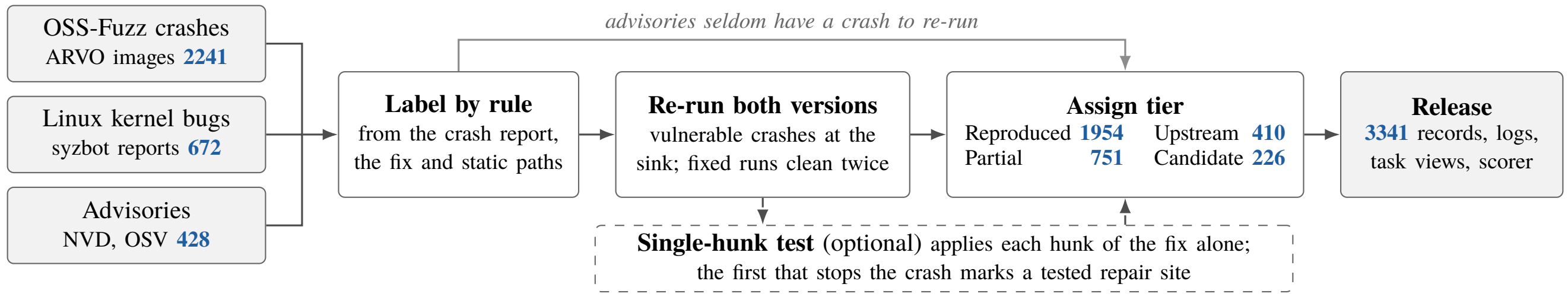

%% file: sections/building.tex
\section{How \CF is built}\label{sec:def}
We take crashes from ARVO and syzbot (Fig.~\ref{fig:pipeline}), label each by
written rules, re-run both versions, and assign a tier. Another \NAdvisoryOnly records come from NVD
and OSV advisories, which seldom have a crash we could run, and none is in the Reproduced tier. Code applies every rule, each dated in a rule log shipped with the data, so no person or model labels a record by hand, and every correction is a logged script change.

\para{Records and labels} A record is an ordered list of CWE \emph{instances} from the root cause to
the sink, each with a CWE, a site (file, line and function) when known, and its own evidence. Most
records have two instances, and \NGoldLenThree in the Reproduced tier add a middle step
that a sanitizer also reported, such as an integer overflow before a write. Each pair of consecutive CWEs must be admissible in our catalog of CWE pairs, and the record
states whether CWE supports it by CanPrecede directly, only through a parent, or not at all (Table S2 in the release's \texttt{docs/SUPPLEMENT.md}). This order does not establish causality or imply separate defects. The sink's CWE follows from the report, so an ASan out-of-bounds read is \cwe{125}, a write \cwe{787}, an MSan uninitialized read \cwe{457}, and, by our rule, a read of freed memory \cwe{125}. Roots have three kinds of site. A \emph{tested repair
site} comes from the single-hunk test. An \emph{origin site} comes from the report. MSan
names where uninitialized memory was allocated, and ASan, MSan or syzbot's KASAN report where used-after-free memory was freed, so
the site marks where bad state began and the report determines the root's CWE (\cwe{456}, missing
initialization, or \cwe{416}, use after free). That site need not be the defect, so since release 1.0.7 a root with a tested site in source code uses it and keeps the origin (Sec.~\ref{sec:val}). A \emph{suggested site} comes from a fix pattern or static analysis.

Each instance carries an evidence category. It is \emph{hunk-named} when one hunk stopped the crash and the check it adds names the CWE, \emph{sanitizer-reported} when a report names it,
\emph{fix-suggested} when the fix adds the check but no run shows it matters, \emph{dataflow}
when only a static path supports it, or \emph{inferred} when only a labeling rule does. Where a root would get \cwe{20}, we use \cwe{1285} (improper validation of an
index) when the fix compares variables like \texttt{idx} or \texttt{offset}, and \cwe{1284} (of a
quantity) for \texttt{len} or \texttt{size}. Of \NGoldRootFormerTwenty such Reproduced-tier roots, \NGoldRootClassLevel fit neither and stay \cwe{20}.

\para{Reproduction and tiers} A run
\emph{crashes} when a sanitizer reports. It is \emph{clean} only when nothing is reported and the
fuzzer's log shows the input was processed (for the kernel, that the reproducer ran its
full time), so a silently failed run never counts as fixed. Kernels are built with syzbot's
configuration and booted in a VM. A record is in the \emph{Reproduced} tier when five
criteria hold. (a)~Our run of the vulnerable version crashes, and the report places the sink.
(b)~The fixed version runs clean twice. (c)~The report agrees with the labels, so the sink's CWE
matches its read or write and no root from static analysis is contradicted, as a buffer overflow contradicts a use-after-free root. (d)~The fix changes non-test code under a recorded commit.
(e)~Any middle step was reported by a sanitizer. This tier certifies the crash, its sink and reproduction, not the root cause. A record is \emph{Upstream} when it meets (d) and (e) and only the reporter's logs show its crash, unless our fixed kernel crashed the same way (\NFixStillCrashes). Our kernel runs reached \NUpstreamRan Upstream records but completed no two-sided run, mostly because the bug did not reproduce (\NUpstreamNoRepro), and their logs were not kept. Every other record our runs reached is \emph{Partial} and lists the criteria it fails. Any remaining record is Partial when its own
material supports the root CWE, such as a dataflow path, a check or lock the fix adds, or the
reporter's sanitizer report. Otherwise it is \emph{Candidate}, since only an advisory's listed CWEs or our catalog's default root for the sink support it. ARVO's fixed image almost always holds other changes (\NFixDistReadyDiff of \NFixDistReadyN checked),
so we record whether the fix alone removes the crash (\emph{fix-attributed}), which for ARVO is the scorer's readiness test and holds by construction for the kernel.

\para{Single-hunk test} We rebuild the vulnerable version and confirm it still crashes, then
apply each non-test hunk of the fix on its own, smallest first, rebuilding and replaying each time.
The first hunk that alone stops the crash marks a tested repair site, and a rule reads the check it
adds to name the root's CWE, such as \cwe{476} (NULL pointer dereference) for a NULL test. A success shows this hunk suffices for this input, not that it is the only or best repair or that the rule's CWE is right. A hunk next to
the crash may only guard the symptom, so each record places its site relative to the crash.

%% file: sections/dataset.tex
\section{The dataset}\label{sec:schema}
\para{Contents} Table~\ref{tab:summary} summarizes \CF. Projects are counted after merging aliases, bugs after grouping records that share a crash, advisory or fix. Single-hunk tests succeeded for \NGoldTestedRoot Reproduced-tier records, and \NRootSiteTested take the tested site as root. For \NGoldProven the hunk's check also named the CWE. For \NGoldTestedOneHunk
(\PctGoldTestedOneHunk) the fix has one code hunk, so the test has only one hunk to try. Common chains go from input-validation failures to out-of-bounds reads (\NChainCheckRead) and writes (\NChainCheckWrite), from missing initialization to uninitialized reads (\NChainUninit) and from use after free to reads (\NChainUafRead), more in Table S2.

\input{tables/summary}

\para{Storage and use} The release is one JSON Lines file (\DatasetMB) with Croissant
metadata~\cite{akhtar2024croissant} and a datasheet~\cite{gebru2021datasheets}, plus \LogsGB of run logs and inputs. Each record's \texttt{chain} lists its instances, and other fields hold
the fix, every run, the pinned images or kernel reproducer, the tier with any
failed criteria, a duplicate group and a split (key fields in Table S5). Field values keep earlier names, such as \texttt{gold} for Reproduced (see the datasheet). The release ships the crashing input of \ArvoInputPhrase, copied from its pinned image. The split keeps each project, and each kernel subsystem, in one of train, validation and test (\NGoldTrain, \NGoldValidation and \NGoldTest Reproduced-tier records), except a catch-all of advisory-only kernel races, and \DupSpanPhrase. Of \NGoldDated dated Reproduced-tier fixes, \NGoldAfterTwentyFive are from 2025 or later, for a temporal holdout. Users filter records, for example to the recommended core, which holds the Reproduced-tier records whose instances were all
hunk-named or sanitizer-reported and whose fix alone removes the crash. They then take each record's task view (Table~\ref{tab:tasks}) and score predictions against the record or with the scorer. Unit tests check that no input leaks a target, though crash reports can reveal one. \CF has no prior published uses.

\input{tables/tasks}

\para{Baselines} Crash reports name the line of
only \NShownTested of \NRootSiteTested tested sites, against \NShownOrigin of \NRootSiteOrigin origins, and the function of \NShownTestedFn tested sites. Tested sites the report reveals stay in evaluation. On the test split, the crashing
function is the root function for \BaseLocFn of \BaseLocN tested sites, and the root function lies on the crash
stack for \BaseLocStack. The modal root CWE per crash type is
right for \BaseRoot of test records (\BaseRootOld before the \cwe{20} mapping, when crash type nearly determined the root CWE).

\para{Patch scorer} The scorer applies a candidate patch to the vulnerable image, rebuilds it, and
replays the input twice. A pass means the input ran to completion with no sanitizer report, not
that the patch is correct. A record is
\emph{ready} when the unmodified build crashes and the developer's fix passes. A patch that just
rejects the input also passes, so for the \NRlCovGold ready records that ship a seed corpus the scorer
also replays the corpus, and its reward can reject a patch that keeps less than \CovTolPct of the unmodified build's
feature coverage. Of \NGoldArvo Reproduced-tier ARVO records,
\NRlScoredGold were scored and \NRlReadyGold are ready, and most others did not apply cleanly or build (Table S3). Kernel records are scored in a VM, where \NRlReadyKernelGold are ready. On GitHub-hosted
machines, the unmodified kernel ran \NKHostedRun of \NKHostedTried reproducers and reproduced the bug in \NKHostedFired.

\para{Where fixes lie} In \PctFixAwayFromCrash of all \NFixAwayBase
Reproduced-tier records, the fix misses the crashing frame's file. In \PctFixOffCrashStackArvo of the \NFixStackBaseArvo Reproduced-tier ARVO records (our kernel KASAN traces lack file names) it touches no file on the crash stack, or \PctFixOffStackArvo counting allocation and free stacks. Of \NTestedRelKnown tested repair sites with a function, \NTestedCrashLine
lie within five lines of the crash and may only guard the symptom, \NTestedCrashFn elsewhere in the crashing function, \NTestedOnStack in a caller and \NTestedOffStack off the stack.

%% file: tables/summary.tex
\begin{table}[t]
\caption{What \CF contains (latest release, v\ReleaseVersion).}
\label{tab:summary}
\centering\footnotesize
\setlength{\tabcolsep}{2.5pt}
\renewcommand{\arraystretch}{1.05}
\begin{tabular}{@{}lr@{\hspace{12pt}}lr@{}}
\toprule
\multicolumn{2}{@{}l@{\hspace{12pt}}}{\textit{All tiers}} &
  \multicolumn{2}{@{}l@{}}{\textit{Reproduced root site}}\\
Records & \NRecords & tested repair site & \NRootSiteTested\\
\quad Reproduced & \NGold & origin site & \NRootSiteOrigin\\
\quad Upstream & \NGoldUpstream & suggested site & \NRootSiteAnnotated\\
\quad Partial & \NSilver & none & \NRootSiteNone\\
\quad Candidate & \NBronze &
  \multicolumn{2}{@{}l@{}}{\textit{Reproduced root evidence}}\\
Projects (with Reproduced) & \NProjects\ (\NProjectsGold) & hunk-named & \NGoldProven\\
\multicolumn{2}{@{}l@{\hspace{12pt}}}{\textit{Reproduced tier}} & sanitizer-reported & \NGoldRootExec\\
Distinct bugs & \NGoldBugs & fix-suggested & \NGoldRootAttested\\
Fix-attributed & \NGoldFixAttr & dataflow & \NGoldRootDataflow\\
Not fix-attributed & \NGoldNotFixAttr & inferred & \NGoldRootInferred\\
Attribution untested & \NGoldFixAttrUnknown & none & \NGoldRootNone\\
Recommended core (projects) & \NCoreFixExec\ (\NCoreProjects) & &\\
\bottomrule
\end{tabular}
\end{table}

%% file: tables/tasks.tex
\begin{table}[t]
\caption{The three tasks.}
\label{tab:tasks}
\centering\footnotesize
\setlength{\tabcolsep}{4pt}
\renewcommand{\arraystretch}{1.1}
\begin{tabularx}{\columnwidth}{@{}>{\bfseries}l >{\hsize=0.94\hsize}Y >{\hsize=1.06\hsize}Y@{}}
\toprule
\normalfont Task & Model input & Target and scoring\\
\midrule
Repair & ARVO image with the fix scrubbed from history; crashing input or kernel reproducer; crash report &
a patch, scored by rebuilding and replaying the input\\
\addlinespace[3pt]
Localize & Repair's input $+$ the sink's CWE and site & tested repair location; file/function agreement where the root is a tested site\\
\addlinespace[3pt]
Label & Repair's input $+$ every instance's site, in order & each instance's CWE, in order\\
\bottomrule
\end{tabularx}\par\vspace{3pt}
\parbox{\columnwidth}{\scriptsize Every view withholds the fix, the fixed image and its runs, the
evidence text, and each label it does not list as input.}
\end{table}

%% file: tables/related.tex
\begin{table*}[t]
\caption{\CF and the closest C/C++ datasets. A dash means none.}
\label{tab:related}
\centering\footnotesize
\setlength{\tabcolsep}{4pt}
\renewcommand{\arraystretch}{1.1}
\newcommand{\no}{\textcolor{black!45}{--}}
\begin{tabular}{@{}l l L{3.75cm} L{2.85cm} L{3.15cm} L{2.55cm}@{}}
\toprule
& & \multicolumn{2}{c}{Labels} & \multicolumn{2}{c@{}}{Execution}\\
\cmidrule(lr){3-4}\cmidrule(l){5-6}
Dataset & Size & What is labeled & Checked by & Evidence & Checked by\\
\midrule
MegaVul~\cite{ni2024megavul} & 17,380 vulns. & unordered advisory CWE IDs & \no & \no & \no\\
PrimeVul~\cite{ding2025primevul} & 6,968 functions & advisory CWE per function & manual sample & \no & \no\\
Magma~\cite{hazimeh2020magma} & 118 bugs (v1.0) & bug type & \no & forward-ported bugs & bug-trigger checks\\
ARVO~\cite{mei2026arvo} & 6,100+ vulns. & crash type; fix commit & patch accuracy (89.4\%) & vulnerable, fixed images & reproduction\\
PatchBench~\cite{shen2026patchbench} & 213 tasks & CWE type; curated patch & manual patch audit & crashing-input replay & input variants, behavior tests\\
\midrule
\textbf{\CF} & \NRecords records & ordered CWEs with evidence; sites where known &
\ifIrrTwoRaters two-rater blind audit\else blind audit in progress\fi; patch-location overlap &
replay logs for executed records; single-hunk tests & our re-runs; re-runs on fresh machines\\
\bottomrule
\end{tabular}
\end{table*}

%% file: sections/validation.tex
\section{Validation}\label{sec:val}
\paraq{Do the crashes reproduce?} Using only the release and its pinned images, GitHub-hosted machines re-ran a
stratified random sample of \RerunCleanN ARVO records that were Reproduced when drawn
(\RerunCleanStillGold still are), and \RerunCleanPct reproduced (95\% CI \RerunCleanCI). They use the same images, so this shows portability, not independent confirmation. Rebuilt from source, \NFromSrcBuilt of \NFromSrcN records built at both versions and \NFromSrcRepro reproduced, so the images are needed. Re-applying \NFuzzTestsAll successful ARVO hunk tests of all tiers with exact context withdrew \NFuzzWithdrawn that applied loosely.

\paraq{Are the root causes right?} A blind audit, whose protocol we dated in the rule log before
recruiting raters, samples \NIrrGold Reproduced-tier and \NIrrOther other records from an earlier version, and the sample and its scored labels ship with the release. Since
then, among other corrections, the \cwe{20} mapping relabeled \NIrrToMapped of their roots, later single-hunk tests and the rule below made \NIrrNowTestedOther suggested and \NIrrNowTestedOrigin origin roots tested repair sites, and \NIrrDropped records were dropped. Raters see the crash report, the
fix, the recorded sites and the dataset's CWE conventions, but not our CWEs or site kinds. They name the CWEs and the root-cause function, and
call each site the defect, a valid repair location, the crash site, unrelated or unclear. The
raters are co-authors who did not build the dataset, and blinding is the safeguard. Each rater worked alone, and no disagreement was settled, so all figures come from the sheets as returned.
\ifIrrTwoRaters
Both are graduate students, the first with a year of secure-code-generation research but little sanitizer experience, the second with C/C++ systems experience but none in security. They agreed on the root CWE
in \IrrAgreeSource of \IrrN items (Cohen's $\kappa$ \IrrKappaSource, 95\% CI \IrrKappaSourceCI; Gwet's
AC1 \IrrAcOneSource) and on the root-site answer in \IrrAgreeSite ($\kappa$ \IrrKappaSite). The site figure compares answers as written and counts \NIrrSiteBothUnclear items both called unclear as agreement. A function-level
comparison added after scoring gives \IrrAgreeSiteFn. Unclear never confirms a label.

Paired figures give the first rater, then the second. On the Reproduced tier the raters agreed with our sink CWE in \IrrSinkCweItems and with our root CWE in \IrrSrcCwePct and \IrrBSrcCwePct. Root agreement rises to \IrrSrcHier and \IrrBSrcHier
when one CWE may specialize the other, and to \IrrSrcHierCur and \IrrBSrcHierCur by that test against today's labels, where a rater's \cwe{20} matches our \cwe{1284} or \cwe{1285}. They judged \IrrPctStrictTested
(95\% CI \IrrStrictTestedCI) and \IrrBPctStrictTested\ (\IrrBStrictTestedCI) of \NIrrTested tested repair sites the defect, but only \IrrPctStrictOrigin\ (\IrrStrictOriginCI) and
\IrrBPctStrictOrigin\ (\IrrBStrictOriginCI) of \NIrrOrigin origin sites and \IrrAnnotPair. They matched our root CWE on only \NIrrTestedCwe and \NIrrBTestedCwe of the tested sites. For origins they matched it in \IrrOriginCwePair but put the defect where the fix edits in \NIrrOriginInFixFn and \NIrrBOriginInFixFn. For the \NIrrMoved origin roots that also had a tested hunk, both in known functions, the raters named the hunk's function in \NIrrMovedTestedFn and \NIrrBMovedTestedFn cases, \NIrrMovedViaCrash and \NIrrBMovedViaCrash of them by choosing the crash site there. Where the two functions differ, they named the origin's in only \NIrrMovedOriginOnly and \NIrrBMovedOriginOnly of \NIrrMovedDiffer. Release 1.0.7 then made such tested sites roots, so these figures describe the rule rather than test it. Raters see the fix, which contains every tested site, and may favor those sites. These rates cover \NIrrLocated Reproduced-tier items with a located root. Weighted by each kind's share of the audited version's \NIrrPop such records with an archived report
(\NIrrPopTested tested, \NIrrPopOrigin origin and \NIrrPopAnnot suggested), the raters judged
\IrrPooledStrict and \IrrBPooledStrict of these roots the defect. Tested repair sites are thus localization targets, not confirmed defects.
\else
So far one rater, a graduate student with about a year of secure-code-generation research who had not read
sanitizer reports in depth, has returned a sheet, so we report that rater's agreement with our
labels, and agreement between raters awaits the second. On the Reproduced tier the rater agreed with
our sink CWE in \IrrSinkCwe and our sink site in \IrrSinkSite. Our root CWE matched in \IrrSrcCwe,
and in \IrrSrcHier one CWE equaled or specialized the other (\IrrSrcHierCur against today's labels, a test that accepts \cwe{20} for \cwe{1284} or \cwe{1285}). The rater judged \IrrStrictTested of
\NIrrTested tested repair sites to be the defect, but only \IrrStrictOrigin of \NIrrOrigin origin
sites and \NIrrStrictAnnot of \NIrrAnnot suggested sites. For origins the CWE was right in \NIrrOriginCwe of
\NIrrOrigin, but the rater put the defect where the fix edits, such as a missing initialization or
an early free, in \NIrrOriginInFixFn cases. Raters see the fix, where tested sites lie, which may favor them. Weighted by each kind's share, \IrrPooledStrict of located Reproduced-tier roots in the
audited version were judged the defect. In an exploratory check planned in the rule log, \NAiModels
language models given the same materials judged \AiTestedRange of tested and \AiOriginRange of origin sites to be
the defect; they are not raters. Tested repair sites are therefore \CF's localization targets, and
an origin site marks where the bad state began.
\fi

\para{Comparison with curated patches} Of Reproduced-tier records, \NPbGold share a fix with a PatchBench bug, whose authors curated a root-cause patch~\cite{shen2026patchbench}. Of these records' \NPbTested tested repair sites, \NPbTestedFn lie in or near a function the patch edits, \NPbTestedFile more
in an edited file and \NPbTestedOther elsewhere. All \NPbOrigin origin sites lie outside its files,
and the \NPbAnnotated suggested sites in or near an edited function. These measure location agreement, not CWE correctness.

%% file: sections/related.tex
\section{Related work}
Table~\ref{tab:related} compares the closest datasets. Magma hand-ports 118 bugs with checks at their sites~\cite{hazimeh2020magma}. ARVO reproduces crashes, and CyberGym and AutoPatchBench build agent tasks on them~\cite{mei2026arvo,wang2026cybergym,meta2025autopatchbench}. MITRE's chains and NIST's Bugs Framework model how weaknesses chain, not executions~\cite{mitre2024chains,nist2024bf}. AURORA and ARCUS perform root-cause analysis~\cite{blazytko2020aurora,yagemann2021arcus}, and Wang et al.\ find replay checks cannot tell root-cause from superficial repairs, which experts can~\cite{wang2026kumushi}. PatchBench curates patches by hand; \CF tests hunks automatically.

%% file: sections/limitations.tex
\section{Research questions, limits and next steps}
\CF supports questions such as these. (1)~How often do repair agents stop the crash, and how often do they edit the tested repair site? A passing replay and matching location do not prove root-cause repair. (2)~Can models predict tested repair sites better than the crashing-function baseline?

ARVO supplies \PctGoldArvo of the Reproduced tier, which inherits OSS-Fuzz's bias toward fuzzable bugs. Of all Reproduced-tier roots, \NGoldUntestedRoot (\PctGoldUntestedRoot) have no
successful single-hunk test, audited origin sites seldom matched the defect, and \ifIrrTwoRaters neither
rater is a security specialist. \else only one audit rater, not a sanitizer expert, has reported so
far. \fi Rule-log dates are self-reported, and each rule from 2026-10-19 on
predates its outcomes unless its entry says otherwise. Replay still needs ARVO's images, pinned by digest but hosted in an unarchived personal Docker Hub account, \texttt{n132/arvo}. Next, we
will \ifIrrTwoRaters\else report agreement between raters, \fi extend hunk-pair tests, use our KASAN runs' free sites as origins and archive the images.

%% file: sections/availability.tex
\section{Availability}
Data, logs, task views, scorer and code are at \DataURL and archived at \DataDOI with Croissant metadata, per the FAIR principles~\cite{wilkinson2016fair}. The README covers setup. We fix errors reported as GitHub issues in versioned releases. In our runs, \ResourceFigures. Code is MIT and labels CC~BY~4.0. Crashing inputs are not under CC~BY~4.0, and diffs, logs and inputs derived from a project's files keep its license, named per record where known. All Reproduced-tier and ARVO bugs are fixed and public, and the only personal data are
developer names and emails the projects publish and our raters' initials and backgrounds.